\documentclass[aps,prd,twocolumn,superscriptaddress,showpacs,amsmath,amssymb,amsthm,nofootinbib, preprintnumbers]{revtex4-2}
 \usepackage{amsmath}
\usepackage{graphicx}
\usepackage{epstopdf}
\usepackage{float}
\usepackage{hyperref}
\usepackage{color}
\usepackage[T1]{fontenc}
\usepackage[utf8]{inputenc}
\usepackage[toc,page]{appendix}
\usepackage[usenames,dvipsnames]{xcolor}
\usepackage[normalem]{ulem}
\usepackage{lipsum, babel}

\newcommand{\tcr}{\textcolor{red}}

\newcommand{\be}{\begin{equation}}
\newcommand{\ee}{\end{equation}}
\newcommand{\ba}{\begin{eqnarray}}
\newcommand{\ea}{\end{eqnarray}}

\newcommand{\beq}{\begin{equation}}
\newcommand{\eeq}{\end{equation}}
\newcommand{\beqa}{\begin{eqnarray}}
\newcommand{\eeqa}{\end{eqnarray}}

\begin{document}

\title{Regularized Conformal Electrodynamics: Novel C-metric in (2+1) Dimensions}

\author{David Kubiz\v n\'ak}
\email{david.kubiznak@matfyz.cuni.cz}
\affiliation{Institute of Theoretical Physics, Faculty of Mathematics and Physics,
Charles University, Prague, V Hole{\v s}ovi{\v c}k{\' a}ch 2, 180 00 Prague 8, Czech Republic}

\author{Otakar Sv{\'i}tek}
\email{ota@matfyz.cz}
\affiliation{Institute of Theoretical Physics, Faculty of Mathematics and Physics,
Charles University, Prague, V Hole{\v s}ovi{\v c}k{\' a}ch 2, 180 00 Prague 8, Czech Republic}

\author{Tayebeh Tahamtan}
\email{tahamtan@utf.mff.cuni.cz}
\affiliation{Institute of Theoretical Physics, Faculty of Mathematics and Physics,
Charles University, Prague, V Hole{\v s}ovi{\v c}k{\' a}ch 2, 180 00 Prague 8, Czech Republic}

\date{April 22, 2024}


\begin{abstract}
Conformal electrodynamics is a particularly interesting  example of power Maxwell non-linear electrodynamics, designed to possess conformal symmetry in all dimensions.   
In this paper, we propose a regularized version of Conformal electrodynamics, minimally regularizing the field of a point charge at the origin by breaking the conformal invariance of the theory with a dimensionfull `Born--Infeld-like' parameter. 
In four dimensions the new theory reduces to the recently studied Regularized Maxwell electrodynamics, distinguished by its `Maxwell-like' solutions for  
accelerated and slowly rotating black hole spacetimes. 
Focusing on three dimensions, we show that the new theory shares many of the properties of its four-dimensional cousin, including the existence of the charged C-metric solution (currently unknown in the Maxwell theory).  
\end{abstract}

\maketitle

\section{Introduction}

{\em Non-linear electrodynamics (NLE)} arose out of attempts to deal with singular nature of classical Maxwell's (linear) theory of electrodynamics when applied to point charges. One of the first and certainly the most famous 
non-linear model was proposed by Born and Infeld almost hundred years ago \cite{Born:1934gh} -- it is distinguished by absence of birefringence and other unique properties \cite{plebanski1970lectures}. Subsequently, other models were proposed to achieve better regularization \cite{Hoffmann:1937noa}, or to embody quantum corrections to Maxwell's theory coming from QED \cite{Heisenberg:1936nmg} and (much later) string theory \cite{Fradkin:1985qd}. More recently, NLE was used as a `physical' source of regular black holes \cite{Ayon-Beato:1998hmi}, increasing its significance for `physics of spacetime'. In this regard, one may formulate new criterion for importance of a given NLE model by demanding its compatibility with essential spacetime geometries (going beyond spherical symmetry) \cite{Tahamtan:2020lvq,Kubiznak:2022vft,Hale:2023dpf, Tahamtan:2023tci}, thus mimicking the success of Maxwell's linear theory in this regard.

While predominantly studied in four dimensions, theories of nonlinear electrodynamics are also interesting in lower/higher-dimensional settings. Among these, {\em Conformal electrodynamics} \cite{Martinez2007} is of particular interest. It is a special example of power Maxwell electrodynamics \cite{PM-Martinez2008}, designed in a way to preserve Weyl symmetry in any number of dimensions, such that in four dimensions it reduces to the Maxwell theory and yields dimension-independent (four-dimensional) Coulomb law for a point charge. 

{In this paper, we propose a {\em regularized version} of Conformal electrodynamics. 
Namely, we design a 1-parametric generalization of Conformal electrodynamics characterized by a dimensionalfull Born--Infeld-like parameter $\alpha$, which yields a finite (minimally regularized) field of a point charge in the origin. While the new {\em Regularized Conformal} electrodynamics naturally breaks the Weyl symmetry of the original theory, it possesses a number of  interesting properties. Namely, in four dimensions it reduces to the recently studied {\em Regularized Maxwell (RegMax)} electrodynamics, which is a unique NLE (constructed from a single field invariant $F_{\mu\nu}F^{\mu\nu}$ that admits `Maxwell-like' Robinson--Trautman \cite{Tahamtan:2020lvq, Tahamtan:2023tci}, C-metric \cite{Hale:2023dpf}, and slowly-rotating \cite{Kubiznak:2022vft} spacetimes (see also \cite{Hale:2024lzh} for a recent discussion of optical properties of the corresponding RegMax black holes). As we shall show in this paper, in three dimensions the regularized theory admits a well behaved generalized charged BTZ black hole with improved thermodynamic charges that are not `plagued' by at infinity logarithmically divergent vector potential. Perhaps most importantly, it also admits a novel charged C-metric solution  (at the moment unknown to exist in 3-dimensional Einstein--Maxwell theory). We shall argue that the last property is very exceptional among all 3-dimensional theories of NLE. 
}

Our paper is organized as follows. The basic properties of NLE theories are reviewed in the next section. Conformal electrodynamics together with an overview of its spherical solutions are gathered in Sec.~\ref{sec3}. The novel Regularized Conformal electrodynamics is proposed in Sec.~\ref{sec4}. Focusing on three dimensions, the  corresponding generalized charged BTZ black holes solutions are studied in Sec.~\ref{sec5}. The novel charged C-metric in $(2+1)$ dimensions is constructed in Sec.~\ref{sec6}. We conclude in Sec.~\ref{sec7}. Appendix~\ref{AppA} overviews spherical charged black holes in the Maxwell theory, while Appendix~\ref{AppB} is devoted to construction of rotating charged BTZ black holes. 


\section{Theories of nonlinear electrodynamics}
\label{sec2}

In this paper we consider Einstein gravity coupled to non-linear electrodynamics, described by the following $d$-dimensional action: 
\begin{equation}\label{bulkAct}
    I= \frac{1}{16\pi} \int_{M} d^dx \sqrt{-g}\left(R +4{\cal L}-2\Lambda\right)\,,
\end{equation}
allowing for a possibility of (negative) cosmological constant $\Lambda$, which we parameterize in terms of the corresponding AdS radius $\ell$ as follows:
\be 
\Lambda=-\frac{(d-1)(d-2)}{2\ell^2}\,,
\ee 
and relate it to the thermodynamic pressure according to, e.g. \cite{Kubiznak:2016qmn}:
\be 
P=-\frac{\Lambda}{8\pi}\,.
\ee 
Here, ${\cal L}$ is the electromagnetic Lagrangian, which is taken to be a function of  electromagnetic field strength invariants of the Maxwell tensor $F_{\mu\nu}=\partial_\mu A_\nu-\partial_\nu A_\mu$ (not considering its covariant derivatives). In $d$ number of spacetimes dimensions, there are up to 
$[d/2]$ such invariants, related to the $[d/2]$ non-trivial eigenvalues of $F_{\mu\nu}$. One convenient way for extracting such eigenvalues is for example to consider  
the traces of the even powers of the Maxwell tensor, namely 
\be 
\mbox{Tr}(F^2)\,,\quad \mbox{Tr}(F^4)\,,\quad \dots \quad \mbox{Tr}(F^{2[d/2]})\,,
\ee
see e.g. \cite{Liu:2019rib} for a construction of quasitopological electromagnetism in terms of powers of such traces. 

A canonical example of 
non-linear electrodynamics is the Born-Infeld theory  \cite{Born:1934gh}, whose Lagrangian in all dimensions is naturally written as (see e.g. \cite{Li:2016nll} for examples of solutions in higher dimensions):
\be 
{\cal L}_{\mbox{\tiny BI}}=-\frac{b^2}{\sqrt{-g}}\sqrt{-\det\Bigl(g_{\mu\nu}+\frac{F_{\mu\nu}}{b}\Bigr)}+b^2\,,
\ee 
where 
$b$ is the Born--Infeld dimensionfull parameter (with dimensions $1/L$), which regularizes the field of a point charge and determines the maximal field strength allowed in the theory. Theories considered in this paper will possess similar parameter.

In this paper we focus on a `simple' class of non-linear theories that are characterized by a single electromagnetic invariant:\footnote{In $d=3$ spacetime dimensions (of main interest in this paper) this is really no restriction, as any NLE therein is characterized by a single field invariant.}
\be\label{POC} 
{\cal L}={\cal L}({\cal S})\,,\quad {\cal S}=\frac{1}{2}F_{\mu\nu}F^{\mu\nu}\,.
\ee
To further restrict the possibilities, one might require that a given theory of non-linear electrodynamics 
should approach that of Maxwell
in the weak field approximation:
\be \label{POC}
\lim_{{\cal S}\to 0}{\cal L}= {\cal L}^{(\mbox{\tiny M})} +o({\cal S})\,,
\quad {\cal L}^{(\mbox{\tiny M})} = -\frac{1}{2} {\cal S}\,,
\ee
a condition known as the
{\em principle of correspondence}.
However, while such a condition is important in four dimensions, there is no reason a priori to consider it in other dimensions as well. In particular, theories studied in this paper will obey the principle of correspondence in four dimensions but will not approach Maxwell's theory in other dimensions.\footnote{Another criterion for restricting possible non-linear theories of electrodynamics is related to the birefringence phenomena, causality, and energy conditions. In this work we shall not deal with these issues and refer the interested reader to recent papers on this topic \cite{Russo:2022qvz, Russo:2024llm, Russo:2024xnh}.}

Introducing the following notation:
\be
{\cal L}_{,{\cal S}}=\frac{\partial {\cal L}}{\partial {\cal S}}\,,\quad {\cal L}_{,{\cal SS}}=\frac{\partial^2 {\cal L}}{\partial {\cal S}^2}\,,
\ee
the {\em generalized Maxwell} equations read
\be\label{FE}
d*D=0\,,\quad
dF=0\,, 
\ee
where 
\be\label{Edef}
D_{\mu\nu} = \frac{\partial \mathcal{L}}{\partial F^{\mu\nu}}
=2{{\cal L}_{,{\cal S}}}F_{\mu\nu}\,. 
\ee
We also obtain the following 
{\em Einstein equations}: 
\be \label{Hmunu}
G_{\mu\nu}+\Lambda g_{\mu\nu}=8\pi T_{\mu\nu}\,,
\ee
where the generalized EM energy-momentum tensor reads 
\be\label{Tmunu}
T^{\mu\nu}=-\frac{1}{4\pi}\Bigl(2F^{\mu\sigma}F^{\nu}{}_\sigma {\cal L_{,S}}-{\cal L}g^{\mu\nu}\Bigr)\,.
\ee
We shall discuss various examples of non-linear electrodynamics below.

\section{Conformal electrodynamics}\label{sec3}

The conformal electrodynamics \cite{Martinez2007} is described by the following Lagrangian: 
\be \label{Conf}
{\cal L}^{(\mbox{\tiny C})}=\frac{2}{d}\beta^{4-d}(-{\cal S})^{d/4}\,,
\ee 
where $\beta$ is a dimensionfull coupling constant, with dimensions $1/\sqrt{L}$. Obviously, in $d=4$ this reduces to the Maxwell electrodynamics \eqref{POC}. Moreover, upon the Weyl scaling $g_{\mu\nu}\to \Omega^2 g_{\mu\nu}$ and $A_{\mu}\to A_{\mu}$, we find that $\sqrt{-g}{\cal L}^{(\mbox{\tiny C})}$ remains in any number of dimensions invariant. We also have 
\be\label{LSC} 
{\cal L}^{(\mbox{\tiny C})}_{,\cal S}=-\frac{1}{2}\beta^{4-d}(-{\cal S})^{\frac{d}{4}-1}=\frac{d}{4{\cal S}}{\cal L}^{\mbox{\tiny (C)}} \,.
\ee 
With this it is easy to check that the corresponding energy-momentum tensor \eqref{Tmunu} is traceless,  $T=T^\mu{}_\mu=0$.

\subsection{Spherical solutions}

In any number of spacetime dimensions, the field of a point charge in conformal electrodynamics is 
given by the (four-dimensional) Coulomb law:
\be 
A_{\mbox{\tiny C}}=-\frac{e}{r}dt\,,
\ee 
where $e$ is a charge parameter of dimensions of length; in what follows (and to simplify our notations) we restrict ourselves to positive charges, $e>0$. Then, $e$ is related to the electric charge according to the following formula:
\be
Q=\frac{1}{4\pi }\int_{S^{d-2}} *D=\frac{\omega_{d-2}}{4\pi} e^{\frac{d-2}{2}}\beta^{4-d}\,,
\ee
where $\omega_{d}$ is the volume of the $d$-dimensional sphere, namely 
\be
\omega_d=\frac{2 \pi^{(d+1)/2}}{\Gamma((d+1)/2)}\,.
\ee

The corresponding spherically symmetric solution is then given by \cite{Hassaine:2007py, Hassaine:2008pw}
(see Appendix~\ref{AppA} for comparison to solutions in standard Einstein--Maxwell theory):
\ba
ds^2&=&-f_{\mbox{\tiny C}} dt^2+\frac{dr^2}{f_{\mbox{\tiny C}}}+r^2 d\Omega_{d-2}^2\,,\nonumber\\
A_{\mbox{\tiny C}}&=&-\frac{e}{r}dt\,,
\ea 
where  $d\Omega_{d}^2$ 
stands for the standard element on $S^d$, and the metric function $f_{\mbox{\tiny C}}$ reads 
\be 
f_{\mbox{\tiny C}}=1-\frac{m}{r^{d-3}}+\frac{4}{d}\frac{e^{d/2}\beta^{4-d}}{r^{d-2}}+\frac{r^2}{\ell^2}\,.
\ee 

One can show that when the above solution describes a black hole, the corresponding thermodynamic quantities are given by (see also \cite{Gonzalez:2009nn, Hendi:2012um}):
\ba
M&=&\frac{d-2}{16\pi}\omega_{d-2}m\,,\quad 
T=\frac{f_{\mbox{\tiny C}}'(r_+)}{4\pi}\,,\quad S=\frac{\omega_{d-2}r_+^{d-2}}{4}\,,\nonumber\\
\phi&=&\frac{e}{r_+}\,,\quad 
V=\frac{\omega_{d-2}r_+^{d-1}}{d-1}\,,\quad 
P=\frac{(d-1)(d-2)}{16\pi \ell^2}\,,\nonumber\\
\Pi_\beta&=&\frac{(d-4)\omega_{d-2}e^{d/2}\beta^{3-d}}{2\pi d r_+}\,,
\ea 
With these at hand, it is easy to verify that the following extended first law holds:
\be\label{First1} 
\delta M=T\delta S+\phi \delta Q+V\delta P+\Pi_\beta \delta \beta\,,
\ee 
which reduces to the standard first law upon fixing the cosmological constant $\Lambda$ and the coupling constant $\beta$. The above first law is accompanied by the corresponding extended Smarr relation, which reads 
\be \label{Smarr1}
(d-3)M=(d-2)TS+(d-3)\phi Q-2PV-\frac{1}{2}\Pi_\beta \beta\,,
\ee 
with the two related by Euler's scaling argument.

\subsection{Conformally charged BTZ black hole}

\begin{figure}
	\begin{center}
		\includegraphics[scale=0.55]{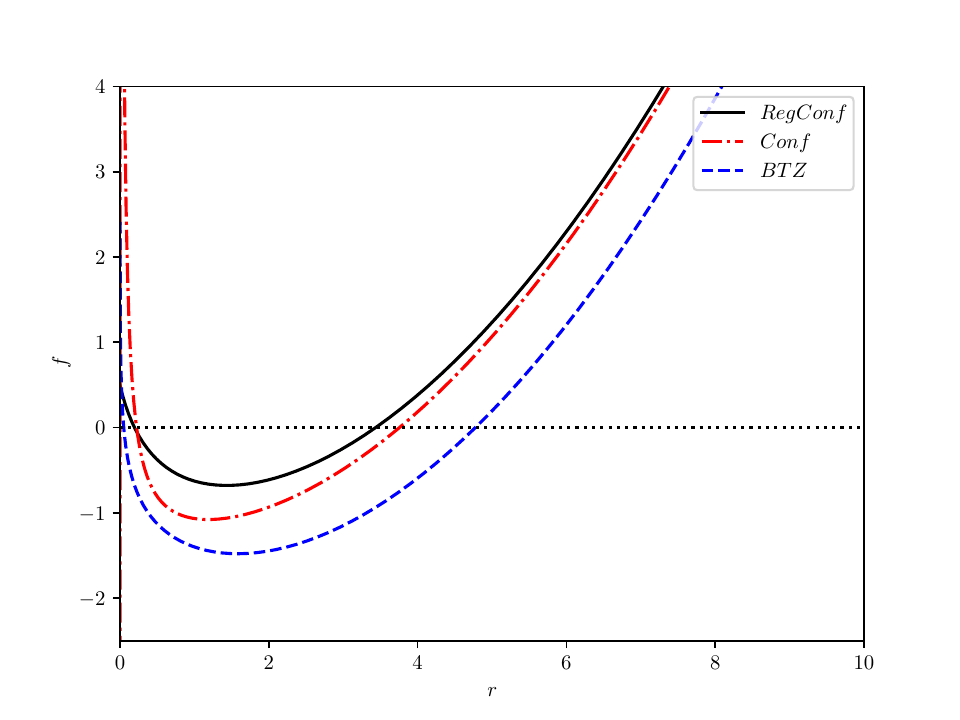}
		\caption{\textbf{Charged BTZ-like black holes.} Here we display the behavior of the metric function $f$ (determining the horizons) for charged BTZ-like black holes in various theories. Namely, the solid black curve corresponds to the Reissner--Nordstr{\"o}m-like RegConf black hole, red dashed-dot curve to a charged black hole in Conformal electrodynamics, and blue dashed  to the charged BTZ black hole in Maxwell's theory. 
  The figure is displayed for $e=1$, $\beta=1$, $\Lambda=-0.1$ and $\alpha=1$. 
		}\label{Fig1}
	\end{center}
\end{figure}

Contrary to Maxwell's case (see Appendix~\ref{AppA}), many of the above formulae remain also valid in $d=3$ dimensions. Let us state these explicitly for future reference. Namely, in $d=3$, the 
conformal electrodynamics reduces to 
\be 
{\cal L}^{(\mbox{\tiny C})}_{3}=\frac{2}{3}\beta(-{\cal S})^{3/4}\,,
\ee 
and admits the following charged BTZ black hole solution 
\cite{Cataldo:2000we, Gurtug:2010dr, Cataldo:2020cxm}:
\ba\label{ConfBTZ}
ds^2&=&-f_{\mbox{\tiny C}}dt^2+\frac{dr^2}{f_{\mbox{\tiny C}}}+r^2d\varphi^2\,,\nonumber\\
A_{\mbox{\tiny C}}&=&-\frac{e}{r}dt\,,
\ea 
where the metric function reads  
\be \label{fCM}
f_{\mbox{\tiny C}}=-m+\frac{4\beta e^{3/2}}{3r}+\frac{r^2}{\ell^2}\,.
\ee 
It demonstrates a typical `Reissner--Nordstr{\"o}m-AdS'-like behavior with two, one extremal, or no black hole horizons. In particular, in Fig.~\ref{Fig1} we display and example of a black hole with two horizons and compare it to other charged BTZ black holes studied in this paper.

The solution is characterized by the following thermodynamic charges: 
\ba
Q&=&\frac{\sqrt{e}\beta}{2}\,,\quad \phi=\frac{e}{r_+}\,,\quad 
M=\frac{m}{8}\,,\nonumber\\
S&=&\frac{\pi r_+}{2}\,,
\quad T=\frac{f'_{\mbox{\tiny C}}(r_+)}{4\pi}=
\frac{r_+}{2\pi \ell^2}-\frac{\beta e^{3/2}}{3\pi r_+^2}\,,\nonumber\\
P&=&\frac{1}{8 \pi \ell^2}\,,\quad V=\pi r_+^2\,,\quad \Pi_\beta=-\frac{e^{3/2}}{3r_+}\,.
\ea
Note that contrary to what happens in Maxwell's electrodynamics, c.f. \eqref{BTZMaxTDs}, thermodynamic volume $V$ here is the `standard' 2d geometric volume.
One can then easily verify that the above thermodynamic quantities obey the generalized first law \eqref{First1} and the Smarr formula \eqref{Smarr1}, which now reduces to a simple relation:
\be 
TS=2PV+\frac{1}{2}\Pi_\beta \beta\,,
\ee 
without explicit $M$ and $\phi Q$ terms.



\section{Regularized conformal electrodynamics}
\label{sec4}
\subsection{Constructing the theory}

Let us now construct a theory which minimally regularizes the conformal electrodynamics.
More precisely, we seek a theory whose vector potential of a pointlike charge in flat space (written in spherical coordinates) takes the following minimally regularized form in any number of dimensions:
\be \label{RegConfA}
A_{\mbox{\tiny RC}}=-\frac{e}{r+r_0}dt\,,\quad r_0=\frac{\sqrt{e}}{
	\alpha}\,.
\ee	
Here we have introduced a dimensionfull `Born--Infeld-like' parameter $\alpha$, which plays the role of a `maximum field strength' and has dimensions $1/\sqrt{L}$; the conformal electrodynamics is recovered upon setting 
\be 
\alpha\to \infty\,.
\ee 
Calculating the field invariant for the above field, we find
\be\label{Suse}
{\cal S}=-\frac{\alpha^4 e^2}{(\alpha r+\sqrt{e})^4}\,. 
\ee
The generalized Maxwell equation   \eqref{FE} in $d$ dimensions then reads  
\begin{equation}
\bigl({\cal L}_{\cal S}\,\frac{\alpha^2 e}{(\alpha r+\sqrt{e})^2}\,r^{d-2}\bigr)_{,r}=0\,.
\end{equation}
Upon integrating this equation and expressing $r$ in terms of ${\cal S}$ via \eqref{Suse}, we recover  
\begin{equation}\label{LSpom}
{\cal L}_{\cal S}=\frac{cs^{d-4}}{(1-s)^{d-2}}\,,
\end{equation}
where $c$ is some (rescaled)  integration constant, and we introduced a shorthand
\be 
s=\Bigl(-\frac{\cal S}{\alpha^4}\Bigr)^{\frac{1}{4}}\in (0,1)\,.
\ee 
Expanding \eqref{LSpom} for large $\alpha$ and comparing it to \eqref{LSC}, fixes the integration constant to $c=-\frac{1}{2}\beta^{4-d}\alpha^{d-4}$, giving 
\be 
{\cal L}_{\cal S}=-\frac{1}{2}\beta^{4-d}\alpha^{d-4}\frac{s^{d-4}}{(1-s)^{d-2}}\,.
\ee 
The full Lagrangian is then obtained by integration:
\be 
{\cal L}=-4\alpha^4\int s^3{\cal L}_{{\cal S}} ds\,.
\ee 
This yields the following {\em Regularized Conformal (RegConf) theory}:
\be
{\cal L}=\frac{2\alpha^d\beta^{4-d}s^d}{d}\mbox{Hypergeom}([d,d-2],[1+d],s)+\mbox{const.}\,,
\ee 
where the integration constant needs to be fixed so that we recover the conformal electrodynamics in the large $\alpha$ limit.


In particular, in the lowest dimensions $d=3, 4, 5, 6$, we recover 
\ba
{\cal L}_3&=&-\alpha^3 \beta \Bigl(s^2+2s+2\log(1-s)\Bigr)\,,\nonumber\\
{\cal L}_4&=&-\alpha^4\Bigl(\frac{s^3+3s^2-6s}{1-s}-6\log(1-s)\Bigr)\,,\nonumber\\
{\cal L}_5&=&-\frac{\alpha^5}{\beta}\Bigl(\frac{s^4+4s^3-18s^2+12s}{(1-s)^2}+12\log(1-s)\Bigr)\,,\nonumber\\
{\cal L}_6&=&-\frac{\alpha^6}{\beta^2}\Bigl(\frac{s^5+5s^4-110s^3/3+50s^2-20s}{(1-s)^3}\nonumber\\
&&\qquad\quad-20 \log(1-s)\Bigr)\,.
\ea
Obviously, ${\cal L}_4$ is nothing else than the Regularized Maxwell Lagrangian studied in \cite{Tahamtan:2020lvq, Kubiznak:2022vft, Hale:2023dpf, Tahamtan:2023tci, Hale:2024lzh}.


\subsection{Three dimensions}
In what follows we shall focus on the regularized conformal electrodynamics in $d=3$ dimensions. Let us summarize here the corresponding formulae. The theory is described by the following Lagrangian:
\ba
{\cal L}&=&-2\beta \alpha^3\,\Bigl(s+\frac{{s}^2}{2}+\ln(1-s)\Bigr)\,,\label{Tay}\qquad\nonumber\\
s&\equiv&\Bigl(-\frac{\mathcal{S}}{\alpha^4}\Bigr)^\frac{1}{4}\in (0,1)\,.\label{s}
\ea
In addition to the dimensionfull parameter $\beta$ of the conformal electrodynamics, the theory is characterized by a new dimensionfull parameter $\alpha>0$, $[\alpha^2]=(\mbox{length})^{-1}$, and reduces to the conformal electrodynamics  in $(2+1)$ dimension upon setting 
 \be
 \alpha\to \infty\,. 
 \ee
 Namely, we have 
 \be\label{POCMadMax}
 {\cal L}=\frac{2}{3}\beta({-\cal S})^{\frac{3}{4}}-\frac{1}{2}\beta\frac{{\cal S}}{\alpha}+
 \frac{2\beta (-{\cal S})^{5/4}}{5\alpha^2}+O(\alpha^{-3})\,;
 \ee
 the limit $\alpha\to 0$ yields the vacuum case. 
 The first and second derivatives of ${\cal L}$ with respect to ${\cal S}$, 
that are important for the field equations and the optical metric, are given by
 \be
 {\cal L}_{\cal S}=-\frac{1}{2}\frac{\beta}{\alpha}\frac{1}{s\,(1-s)}\,, \quad 
 {\cal L}_{\cal SS}=\frac{\beta}{8 \alpha^5}\frac{(2s-1)}{s^{5}\,(s-1)^2}\,,
 \ee
We shall now turn to constructing simple (black hole) solutions in this theory.

\section{Generalized charged BTZ black hole}\label{sec5}
Let us first show that the Regularized Conformal electrodynamics in $d=3$ dimensions admits a charged BTZ-like black hole solution, generalizing \eqref{ConfBTZ}. 
It takes the following simple form:
\ba\label{SSS}
ds^2&=&-f_{\mbox{\tiny RC}}dt^2+\frac{dr^2}{f_{\mbox{\tiny RC}}}+r^2d\varphi^2\,,\\
A_{\mbox{\tiny RC}}&=&-\frac{\alpha e}{\alpha r+\sqrt{e}}dt\,,
\ea
where the metric function $f_{\mbox{\tiny RC}}$ reads 
\ba 
f_{\mbox{\tiny RC}}&=&2\alpha\beta e-m-4\sqrt{e}\alpha^2\beta r+\frac{r^2}{\ell^2}\nonumber\\
&&+
4\alpha^3\beta r^2\log\Big(\frac{\alpha r +\sqrt{e}}{r\alpha}\Bigr)\,.
\ea


\begin{figure}
	\begin{center}
		\includegraphics[scale=0.55]{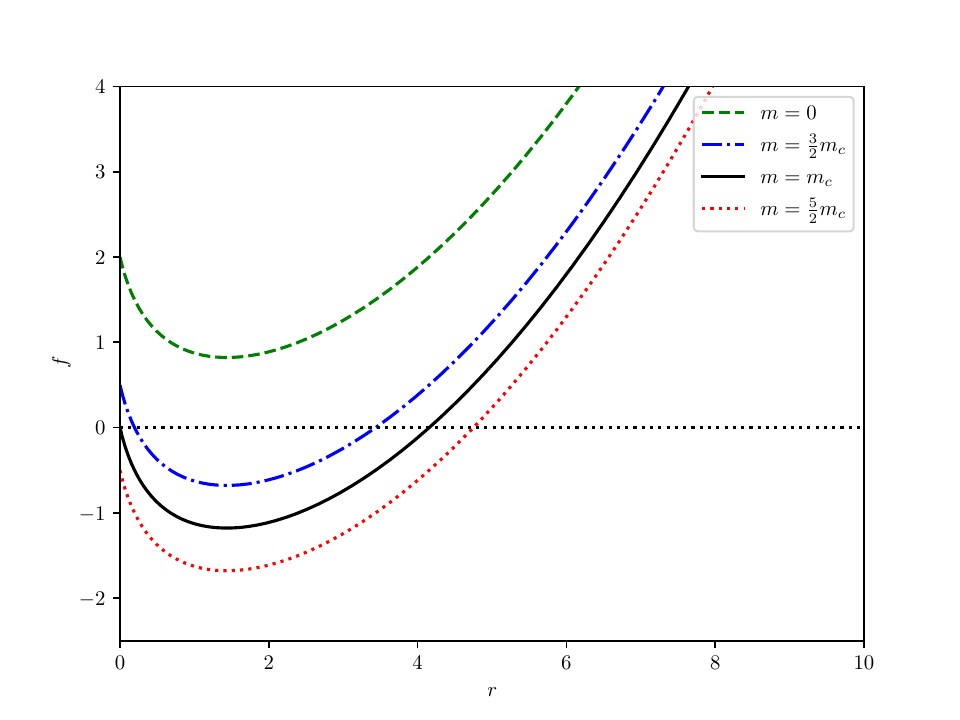}
		\caption{\textbf{Three types of RegConf black holes.} Based on the bahavior of the metric function $f_{\mbox{\tiny RC}}$ near the origin, we distinguish three types of black holes:  i) Reissner--Nordstr{\"o}m branch (dot-dashed blue) ii) Schwarzschild branch (dotted red), and iii)  marginal case $m=m_c$ (solid black). This is also compared to a (naked singularity) $m=0$ solution in the Reissner--Nordstr{\"o}m branch (green dashed). The figure is displayed for $e=1$, $\beta=1$, $\Lambda=-0.1$ and $\alpha=1$.  
		}\label{Fig2}
	\end{center}
\end{figure}

The corresponding field strength
\be\label{E}
F=dA_{\mbox{\tiny RC}}=Edr\wedge dt\,, \quad {E}=\frac{e\alpha^2}{(\alpha r+\sqrt{|e|})^2}\,,
\ee
approaches a finite value in the origin, 
$E_0=E(r=0)=\alpha^2$.
For large $\alpha$ (or alternatively large $r$) we recover the conformal electrodynamics metric function:
\be 
f_{\mbox{\tiny RC}}= 
f_{\mbox{\tiny C}}-{\frac {\beta{e}^{2}}{r^{2}\alpha}}+O(1/\alpha^2)\,.
\ee
On the other hand, near the origin,  $r\to 0$, we find 
\be 
f_{\mbox{\tiny RC}} =2\alpha \beta e-m-4\sqrt{e}\alpha^2\beta r+O(r^2)\,.
\ee
Although the metric function remains finite at the origin, the black hole solution possesses a singularity at $r=0$, as can for example be seen by expanding the Ricci scalar. In particular, setting $\Lambda=0$ for the moment, we find the following expansions for the Ricci and Kretschmann scalars:
\ba 
R&=&-\frac{8\sqrt{e}\alpha^2\beta}{r}+O(\log r)\,,\nonumber\\
{\cal K}&=&R_{\alpha\beta\gamma\delta}R^{\alpha\beta\gamma\delta}=\frac{32 e\alpha^4\beta}{r^2}+O\Bigl(\frac{\log r}{r}\Bigr)\,.
\ea 
This is to be compared to the vanishing Ricci scalar of the conformal BTZ black hole \eqref{ConfBTZ}, as well as to its (significantly more divergent) Kretschmann scalar, 
${\cal K}_{\mbox{\tiny C}} = \frac{32e^3\beta^2}{3r^6}$.

In order to have a black hole, we have to have $\Lambda<0$. Dependent on the choice of  parameters we then obtain three `types' of black holes, see Fig.~\ref{Fig2}. Namely, 
since at $r=0$ the metric function $f_{\mbox{\tiny RC}}$ remains finite:
\be 
f_{\mbox{\tiny RC}} =m_c-m\,,\quad 
m_c=2\alpha\beta e>0\,,
\ee 
if 
$m<m_c$, we have the {\em `Reissner--Nordstr{\"o}m' branch}, with timelike singularity and two, one extremal, or no black hole horizons. If $m>m_c$, we have the {\em `Schwarzschild' branch}
with spacelike singularity and one black hole horizon. Finally, $m=m_c$ is the {\em marginal case}, characterized by $f_{\mbox{\tiny RC}}(r=0)=0$. Formally, the origin becomes a `horizon', though the curvature scalars  still diverge there. At the same time this `place' is point-like since we use the `area  gauge' for the coordinate $r$. This behavior is similar to what happens in the scalar field spacetimes in four-dimensional general relativity \cite{JNW:1968}, where a notion of the so-called `black point' is used for its description. This type of `null point-like singularity' prevails also for scalar field spacetimes in the presence of non-linear electrodynamics 
\cite{Tahamtan-PRD:2020}.


The above generalized BTZ black hole 
can be assigned the following 
thermodynamic quantities:
\ba
M&=&\frac{m}{8}\,,\quad S=\frac{\pi r_+}{2}\,,\quad V=\pi r_+^2\,,\nonumber\\
T&=&\frac{f_{\mbox{\tiny RC}}'(r_+)}{4\pi}\,, \quad \phi=\frac{\alpha e}{\alpha r_++\sqrt{e}}\,, \quad Q=\frac{\beta \sqrt{e}}{2}\,,
\ea
together with 
\ba 
\Pi_\beta&=&\frac{r_+^2\alpha^3}{2}\log\Bigl(1+\frac{\sqrt{e}}{\alpha r_+}\Bigr)-\frac{e\alpha(2\alpha^2r_+^2+\alpha r_+\sqrt{e}+e)}{4(\alpha r_+\sqrt{e}+e)}\,,\nonumber\\
\Pi_\alpha&=&
\frac{3}{2}r_+^2\alpha^2\beta\log\Bigl(1+\frac{\sqrt{e}}{\alpha r_+}\Bigr)\nonumber\\
&&+\frac{\beta(e^2-9e\alpha^2r_+^2-2e^{3/2}\alpha r_+-6\sqrt{e}\alpha^3r_+^3)}
{4(\alpha^2r_+^2+2\alpha r_+\sqrt{e}+e)}\,.
\ea 
It is easy to verify that these obey the following extended first law:
\be 
\delta M = T\delta S+\phi \delta Q+V\delta P
+\Pi_\alpha \delta \alpha+\Pi_\beta \delta \beta\,,
\ee 
together with the corresponding Smarr relation:
\be 
TS=2PV+\frac{1}{2}\Pi_\alpha \delta \alpha+\frac{1}{2}\Pi_\beta \delta \beta\,.
\ee 
Unfortunately, contrary to their four-dimensional cousins, the 3-dimensional RegConf black holes do not seem to admit  
any remarkable thermodynamic behavior.

So far we have focused on static RegConf black holes. However, the rotating ones can easily be obtained by a `boost trick' a la \cite{Clement:1993kc, Clement:1995zt, PhysRevD.61.104013} -- we present such solutions in Appendix~\ref{AppB}. Here, we focus on a more interesting solution -- a solution describing RegConf accelerated black holes.

\section{Novel charged C-metric in (2+1) dimensions}\label{sec6}
Recently there has been a lot of interest in 3-dimensional C-metric \cite{3dCmetric2011, 3dCmetric2012}, see e.g. \cite{3dCmetric2022, Fontana:2024odl} for analysis of the solution and \cite{Arenas-Henriquez:2023hur, Tian:2024mew, Tian:2023ine} for attempts at its thermodynamic interpretation. Since its first discovery, the vacuum solution \cite{3dCmetric2011, 3dCmetric2012} has been generalized to include scalar field \cite{Cisterna:2023qhh} and most recently also to conformal electrodynamics \cite{3dCmetric2023}. While the solution may not exist in Maxwell's theory (see below), in this section we generalize it to the Regularized Conformal electrodynamics.

\subsection{Accelerated BTZ black hole}
The $(2+1)$-dimensional vacuum C-metric is most easily written in the so called `$x-y$' coordinates \cite{3dCmetric2011, 3dCmetric2012, 3dCmetric2022}, and reads 
\be\label{XYansatz}
ds^2=\frac{1}{\Omega^2\!(x,y)}\Bigl(-F(y)dt^2+\frac{dy^2}{F(y)}+\frac{dx^2}{G(x)}\Bigr)\,, 
\ee
where the conformal factor is 
\be \label{OmegaBTZ}
\Omega=g(x+y)\,,
\ee 
with $g$ an acceleration parameter (of dimensions of $1/L$), and the metric functions $F$ and $G$ that take the following form:
\ba 
F&=&\frac{1}{2}c_1 y^2+c_2y+c_3+\frac{1}{g^2\ell^2}\,,\nonumber\\
G&=&-\frac{1}{2}c_1 x^2+c_2x-c_3\,.
\ea 
Such a metric can describe accelerated particle-like solutions or black holes.  
Focusing on accelerated black holes that are smoothly connected to a BTZ black hole, we set 
\be 
c_1=-2\,,\quad c_2=0\,,\quad c_3=1\,,
\ee 
upon which 
\be 
G=x^2-1\,,\quad F=1-y^2+\frac{1}{g^2\ell^2}\,,
\ee 
and (to preserve the signature of the spacetime) we have to have $x>1$ (or alternatively $x<-1$). More concretely, 
let $x\in (x_{\mbox{\tiny min}}, x_{\mbox{\tiny max}})$ be the appropriate range of coordinate $x$, with $x_{\mbox{\tiny min}}>1$. The zeroes of the metric function $F$ determine the position of the black hole ($y_+$) and Rindler ($y_R$) horizons; explicitly these are given by 
\be 
y_+=-\frac{\sqrt{1+g^2\ell^2}}{g\ell}\,,\quad 
y_R=\frac{\sqrt{1+g^2\ell^2}}{g\ell}\,.
\ee 
We can now distinguish two cases: i) the case of {\em rapid acceleration}, which happens for $x_{\mbox{\tiny max}}>y_R$, and in which case both horizons are present and ii) the case of {\em slowly accelerating} black holes with no Rindler horizon, for which $x_{\mbox{\tiny max}}<y_R$.

To make connections with the 4-dimensional C-metric, e.g. \cite{Anabalon:2018ydc, Anabalon:2018qfv}, let us perform the following change of coordinates \cite{3dCmetric2022}:
\be 
r=-\frac{1}{{\cal A}y}\,,\quad x=\cosh(m\phi)\,,\quad t=\frac{m^2{\cal A}\tau}{\omega}\,,\quad {\cal A}=\frac{g}{m}\,,
\ee 
upon which the metric takes a `more familiar form':
\be 
ds^2=\frac{1}{\Omega^2}\Bigl(-f\frac{d\tau^2}{\omega^2}+\frac{dr^2}{f}+r^2d\phi^2\Bigr)\,,
\ee 
where 
\ba
f&=&\frac{r^2}{\ell^2}-m^2(1-{\cal A}^2r^2)\,,\nonumber\\
\Omega&=&1+{\cal A}r \cosh(m\phi)\,,
\ea 
Here, $m=\mbox{arcosh}(x_{\mbox{\tiny max}})/\pi$ regulates the tension of the wall pulling the black hole and ensures that $\phi\in (-\pi,\pi)$, and $\omega$ is not an independent parameter, but rather a combination of other parameters, ensuring the proper normalization of the proper time of an asymptotic observer.

We shall not attempt to review more properties of the above solution here and refer the interested reader to the original literature above. Instead, we proceed directly to finding the corresponding charged generalization in Regularized Conformal electrodynamics.

\subsection{Regularized charged C-metric}
{To find the charged generalization of the above vacuum solution in Regularized Conformal electrodynamics, we employ the ansatz \eqref{XYansatz} and \eqref{OmegaBTZ}, and accompany it with the following ansatz for the vector potential:
\be \label{Aansatz}
A=\psi(y) dt\,.
\ee 
The Einstein equations together with the generalized Maxwell equation then yield the following solution: 
\ba 
F&=&\frac{4\beta \alpha^3}{g^2}\log\Bigl(y+\frac{\alpha}{g\sqrt{e}}\Bigr)+\frac{c_1}{2}y^2+c_2y+c_3+\frac{1}{\ell^2 g^2}\,,\nonumber\\
G&=&-\frac{4\beta \alpha^3}{g^2}\log\Bigl(\frac{\alpha}{g\sqrt{e}}-x\Bigr)-\frac{c_1}{2}x^2+c_2x-c_3\,,
\ea 
and 
\be
A=\frac{\alpha ey}{\alpha+\sqrt{e}gy}dt\,.
\ee 
The solution looks remarkably similar to its four-dimensional cousin in this theory, c.f. Sec.~V.A in \cite{Hale:2023dpf}. 
In particular, note that when the cosmological constant vanishes, the functions $F(y)$ and $G(x)$ have the following property: $F(w)=-G(-w)$. Similar to the vacuum case, to maintain a Lorentzian signature of the metric \eqref{XYansatz}, it is necessary
that $G > 0$, which 
implies restrictions on the domain of coordinate $x$. 
We postpone the detailed discussion of this solution to a future study. Here we only make two remarks.
}


First, 
in the large $\alpha$ limit we recover the charged C-metric in conformal electrodynamics studied in \cite{3dCmetric2023}, namely:
\ba 
F&=&\frac{4}{3}e^{3/2}g\beta y^3+\frac{c_1}{2}y^2+c_2y+c_3+\frac{1}{g^2\ell^2}\,,\nonumber\\
G&=&\frac{4}{3}e^{3/2}g\beta x^3-\frac{c_1}{2}x^2+c_2x-c_3\,,
\ea 
together with 
\be 
A=ey dt\,.
\ee

{Second, one can easily check that while we were able to construct the charged C-metric for the conformal electrodynamics and its regularized generalization, the ansatz \eqref{XYansatz} together with \eqref{Aansatz} are incompatible with many other theories of NLE, including the Maxwell theory. Namely, starting with any NLE, and using the ansatz \eqref{Aansatz}, the time component of the modified Maxwell equation, $(\nabla \cdot D)_t=0$, can be once integrated, to yield   
\begin{equation}\label{psiprime}
	\psi_{,y}=\frac{c(x)}{\Omega(x,y)\,\mathcal{L}_{S}(y,x)}\,,
\end{equation}
where $c=c(x)$ is an integration `constant', a function of $x$-coordinate only. 
However, since the corresponding $T_{\mu\nu}$ obeys $T_{xy}=0$, we must also have $G_{xy}\propto \Omega_{,xy}=0$, that is, $\Omega$ has to be separable:
\be\label{HxHy}
\Omega=\Omega_x(x)+\Omega_y(y)\,.
\ee
Moreover, for the spacetime to describe the C-metric, as we know it, both such parts have to be non-trivial. 
Thus the previous equation \eqref{psiprime} imposes a very strict restriction on the form of ${\cal L}_{\cal S}$ for a given theory, namely 
\be\label{condCmetric} 
{\cal L}_{S}(x,y)=\frac{c(x) h(y)}{\Omega(x,y)}\,,
\ee 
for some function $h=h(y)$. Obviously, for Maxwell, ${\cal L}_{\cal S}=-1/2$ and the previous equation cannot be satisfied. Remarkably, for the Regularized C-metric solution above, we find
\be 
{\cal L}_{S}=\frac{\beta}{2g\alpha\sqrt{e}}\frac{(yg\sqrt{e}+\alpha)^2}{(x+y)(xg\sqrt{e}-\alpha)}\,,
\ee 
which is precisely of the form above. It remains to be seen, whether the Regularized Conformal electrodynamics is the most general theory for which the equation \eqref{condCmetric} 
can be satisfied.  
}

\section{Summary}\label{sec7}
Conformal electrodynamics is a very interesting example of a power Maxwell theory characterized by preserving the Weyl symmetry in any number of dimensions. In four dimensions it coincides with the Maxwell theory, while it breaks the principle of correspondence in any other dimension.  

In this paper we have generalized 
the recently studied four-dimensional RegMax electrodynamics to any number of dimensions. A foundational feature of the new theory (inherited from its four-dimensional cousin) is that it minimally regularizes the field of a point charge -- it is characterized by a dimensionfull Born--Infeld-like parameter, which imposes a maximal bound on the field strength at the position of the charge. In addition, we have designed our theory so that in any number of dimensions it reduces to the 
Conformal electrodynamics (and in particular to the Maxwell theory in four dimensions) in the weak field limit -- thence the name Regularized Conformal 
electrodynamics.

Moreover, focusing on three dimensions, we have shown that the new theory admits charged BTZ-like black holes with vanishing at infinity vector potential, giving rise to a much simpler thermodynamic interpretation than is the case of the Maxwell charged BTZ black holes whose potential logarithmically diverges at infinity. Even more remarkably, the theory admits a 3-dimensional generalization of a charged C-metric, thus providing a non-trivial example of charged accelerating black holes in three dimensions, a property it shares with its four-dimensional RegMax cousin. We suspect that Regularized Conformal electrodynamics may be the most general theory for which the accelerated charged black holes can be found in the above studied form. 

The extension of the RegMax theory beyond four dimensions is of course not unique. For example, instead of demanding that the theory in the weak field limit approaches that of the Conformal electrodynamics, we might have required it to approach the Maxwell electrodynamics instead, giving rise to `genuine' RegMax electrodynamics in all dimensions. Namely, instead of the potential \eqref{RegConfA} we could have demanded that it goes like:
\be 
A=-\sqrt{\frac{d-2}{2(d-3)}}\frac{q}{(r+r_0)^{d-3}}dt\,,\quad {r_0=\frac{q^{\frac{1}{2(d-3)}}}{\alpha}}\,,
\ee 
c.f. Eq.~\eqref{AMmax} in Appendix~\ref{AppA}(or, in $d=3$ dimensions $A=-Q\log(r/r_0+1)$ instead of \eqref{BTZchargedM}. However, the corresponding theory does not seem to give raise to C-metric solutions in three dimensions. It may, however, be interesting in higher dimensions, for example in connection with slowly rotating black holes. We leave this endeavor for future studies.

\section*{Acknowledgements}
 D.K. and T.T. are grateful for support from GA{\v C}R 23-07457S grant of the Czech Science Foundation. O.S. is supported by Research Grant No. GA{\v C}R 22-14791S. D.K. acknowledges the Charles University Research Center Grant No. UNCE24/SCI/016.

\appendix

\section{Charged black hole in Maxwell theory}\label{AppA}

The charged AdS black holes in Maxwell theory in $d>3$ number of spacetime dimensions take the following standard form, e.g. \cite{Gunasekaran:2012dq}: 
\ba\label{AMmax}
ds^2&=&-f_{\mbox{\tiny M}} dt^2+\frac{dr^2}{f_{\mbox{\tiny M}}}+r^2 d\Omega_{d-2}^2\,,\nonumber\\
A_{\mbox{\tiny M}}&=&-\sqrt{\frac{d-2}{2(d-3)}}\frac{q}{r^{d-3}}dt\,,
\ea 
where  $d\Omega_{d}^2$ 
stands for the standard element on $S^d$, and the metric function $f_{\mbox{\tiny M}}$ reads 
\be 
f_{\mbox{\tiny M}}=1-\frac{m}{r^{d-3}}+\frac{q^2}{r^{2(d-3)}}+\frac{r^2}{\ell^2}\,.
\ee 
Here, the electric charge $Q$ is given by
\be 
Q=\frac{\sqrt{2(d-2)(d-3)}}{8\pi} \omega_{d-2}q\,,
\ee
and the mass $M$ reads  
\be 
M=\frac{d-2}{16\pi}\omega_{d-2}m\,. 
\ee 
The remaining thermodynamic quantities are 
\ba
T&=&\frac{f_{\mbox{\tiny M}}'(r_+)}{4\pi}\,,\quad S=\frac{\omega_{d-2}r_+^{d-2}}{4}\,,\nonumber\\
\phi&=&\sqrt{\frac{d-2}{2(d-3)}}\frac{q}{r_+^{d-3}}\,,\quad V=\frac{\omega_{d-2}r_+^{d-1}}{d-1}\,.
\ea 
Together, they obey 
the standard first law of thermodynamics: 
\be \label{firstAppA}
\delta M=T\delta S+\phi \delta Q+V \delta P\,.
\ee

The above solution is valid in $d>3$ dimensions. For $d=3$, one has instead the charged BTZ black hole  \cite{BTZ1992, RotatingBTZ1993}. 
It reads as follows:
\ba \label{BTZchargedM}
ds^2&=&-f_{\mbox{\tiny M}}dt^2+\frac{dr^2}{f_{\mbox{\tiny M}}}+r^2 d\varphi^2\,,\nonumber\\
A_{\mbox{\tiny M}}&=&-Q\log(r/r_0)dt\,,
\ea 
where 
\be 
f_{\mbox{\tiny M}}=-m-\frac{Q^2}{2}\log(r/r_0)+\frac{r^2}{\ell^2}\,.
\ee 
Here, $r_0$ is a dimensionfull constant with dimensions of length; often in the literature $r_0$ is simply associated with the AdS length scale, $r_0=\ell$, e.g. \cite{Frassino:2015oca}. The logarithmic divergence renders calculation of the asymptotic mass a bit problematic, see, e.g.
\cite{Cadoni:2007ck} for a a possible renormalization procedure. In any case, the solution can be assigned the following thermodynamic quantities \cite{Frassino:2015oca}:
\ba\label{BTZMaxTDs} 
M&=&\frac{m}{8}\,,\quad T=\frac{f'_{\mbox{\tiny M}}(r_+)}{4\pi}\,,\quad S=\frac{\pi r_+}{2}\,,\quad P=\frac{1}{8\pi \ell^2}\,,\nonumber\\
\phi&=&-\frac{Q}{8}\log(r/\ell)\,,\quad  
V=\pi r_+^2-\frac{Q^2\pi \ell^2}{4}\,;
\ea 
note the departure of $V$ from the `standard geometric volume'. In any case, the above thermodynamic quantities obey the above standard first law \eqref{firstAppA}.

\section{Rotating charged BTZ black holes}\label{AppB}

Rotating charged BTZ-like black holes (in any NLE) can be obtained from non-rotating ones by the following trick \cite{Clement:1995zt, PhysRevD.61.104013}. Start from a static solution
\ba
ds^2&=&-fdt ^2+\frac{dr^2}{f}+r^2d\varphi^2\,,\nonumber\\
A&=&\psi dt\,,
\ea
and apply the following boost:
\be 
t =\frac{T-\omega\phi}{\sqrt{1-\omega^2}}\,,\quad \varphi =\frac{\phi-\omega T}{\sqrt{1-\omega^2}}\,.
\ee
This, upon the right identification of the new coordinate $\phi$, yields the rotating and charged BTZ-like black hole solution 
\ba
ds^2&=&-N^2FdT^2+\frac{dR^2}{F}+R^2(d \phi+N^{\phi} dT)^2\,,\nonumber\\
A&=&\frac{\psi}{\sqrt{1-\omega^2}}\bigl(dT-\omega d\phi\bigr)\,,
\ea
where
\ba
    R^2&=&\frac{r^2-\omega^2 f}{1-\omega^2}\,,\quad F=\left(\frac{dR}{dr}\right)^2f\,, \nonumber\\
    N^{\phi}&=&-\frac{\omega (r^2-f)}{R^2\,(1-\omega^2)}\,,\quad  N=\frac{r}{R}\left(\frac{dr}{dR}\right)\,. 
\ea


\providecommand{\href}[2]{#2}\begingroup\raggedright\endgroup

\end{document}